# Observation of large anisotropy in *c*-axis oriented thin films of MgB$_2$ prepared by an *in situ* pulsed laser deposition process.

S.R. Shinde[1,*], S.B. Ogale[1,2,#], A. Biswas[1], R.L. Greene[1], and T. Venkatesan[1]

[1]Center for Superconductivity research, Department of Physics, University of Maryland, College Park, MD 20742.
[2]Department of Materials and Nuclear Engineering, University of Maryland, College Park, MD 20742.

**Abstract**:
A large anisotropy in the upper critical field ($H_{C2}$) is observed for MgB$_2$ films grown *in situ* by a pulsed laser deposition process involving growth and annealing of Mg and B multilayers. Measurements of resistivity as a function of temperature and magnetic field yield the estimated zero temperature values of $H_{C2}$ in the range 100 – 130 T and about 10 T, for the field in *ab* plane and along *c*-axis, respectively, depending on the criterion chosen for the transition temperature. The corresponding anisotropy parameter is thus in the range 9 – 13. The estimated coherence length in *ab* plane is about 50 Å, whereas that in the *c*-direction is much smaller (~ 5 Å). No significant magnetoresistance is observed in the normal state.

**Introduction**:

The recent discovery[1] of superconductivity in MgB$_2$ has renewed interest in the area of intermetallic superconductors mainly because (i) it introduced a concept of superconductivity in a new class of materials viz., boride family, (ii) it has much higher transition temperature ($T_C$ ~ 39 K) than other intermetallic superconductors, (iii) unlike high $T_C$ cuprates, the weak link problem due to grain boundaries has been shown to be minimal in these compounds, and (iv) it is commercially available low cost material and can be easily made in the wire form. The replacement of $^{11}$B by its isotope $^{10}$B enhanced $T_C$ from 39.2 to 40.2 K, indicating the phonon-mediated nature of superconductivity.[2] This observation of isotope effect was very soon supported by detecting the BCS-type energy gap by tunneling spectroscopy[3] and $^{11}$B NMR experiments[4]. The higher $T_C$ observed in MgB$_2$ has been attributed to the enhanced phonon frequency due to light ionic masses.

Immediately after the discovery, a large number of researchers have measured several fundamental properties and parameters of MgB$_2$. However, the observations reported by different groups are sometimes drastically dissimilar. For example, Finnemore *et al.*[4] and Bud'ko *et al.*[2] observed large magnetoresistivity in the normal state of sintered polycrystals and dense polycrystalline wires of MgB$_2$ and the measured value of the upper critical field, $H_{C2}(0)$, for both the cases was ~ 16 T, with a slope, $dH_{C2}/dT$, of about 0.44 T/K. On the other hand, no magnetoresistance was observed by Pradhan *et al.*[6] in their wires, and a very small magnetoresistance was reported for *ex situ* thin films[7]. For *ex situ* thin films, Jung *et al.*[7] observed much higher $H_{C2}(0)$, 24 ± 3 T for the magnetic field ($H$) oriented along $c$-axis and 30 ± 2 T for the field in *ab* plane, than normally observed in different forms of polycrystalline materials[2,4]. However, Patnaik *et al.*[8] and Eom *et al.*[9] showed that with substantial amount of oxygen present in the film, the $H_{C2}(0)$ and the irreversibility field, $H^*(0)$, is enhanced to much higher values. For example, for the film with composition Mg:B:O = 1.0 : 0.9 : 0.7, the observed values of $^cH_{C2}(0)$ and $^{ab}H_{C2}(0)$ were about 19.5 T and 39 T, respectively. Here the prefixes "$c$" and "$ab$" indicate that $H$ is along $c$-axis and in *ab* plane, respectively.

Another important parameter reflecting a degree of ambiguity is the anisotropy ratio, $\eta$, defined as the ratio of upper critical fields determined with $H$ in *ab* plane and along $c$-axis, i.e. $^{ab}H_{C2}(0)/^cH_{C2}(0)$. Due to the planer structure of MgB$_2$, one expects a large anisotropy in its magnetic properties. Since the earlier studies were performed on polycrystalline samples, the nature of anisotropy was not revealed. However, an anisotropic superconducting gap was inferred from a number of indirect measurements and was suggested in several theoretical studies.[10] Papavassiliou *et al.*[10] have shown from their NMR data that MgB$_2$ is strongly anisotropic. They observed magnetic field distribution in NMR signal and with complementary use of magnetization data, they deduced $\eta$ ~ 6. Simon *et al.*[12] also observed strong anisotropic nature of MgB$_2$ in their conduction electron spin resonance (CESR) as well as high field reversible magnetization data on powder samples. However, there is no direct observation of such a large anisotropy. The direct measurements done by Lima *et al.*[13] on aligned MgB$_2$ crystallites show that for these samples $\eta$ ~ 1.7 with $^cH_{C2}(0)$ ~ 13 T. After the recent success in growing sub-millimeter size MgB$_2$ single crystals, the resistivity measurements performed by Lee et al.[14] show that $\eta \sim 2.7$, with $^cH_{C2}(0) \sim 9 - 10$ T. Somewhat smaller anisotropy has also been observed in c-axis oriented thin films prepared by *ex situ* approach[7,8]. The values of $\eta$ reported for thin films are 1.24 [Ref. 7] and $1.8 - 2$ [Ref. 8].

Since $MgB_2$ is a highly promising candidate for technological applications, it is important to resolve the issue of magnitude of anisotropy in this system. Because the reports on different forms of $MgB_2$ are drastically different, it is useful to examine the anisotropic nature of *in situ* films, which are expected to play a key role in superconducting electronic applications and sensor technology. To the best of our knowledge, there is no report of the anisotropy in thin films prepared by an *all in situ* approach. In this paper, we report the results of resistivity measurements in magnetic field up to 8 T applied along c-axis and in the *ab* plane. The results presented here reveal anisotropy parameter $\eta$ in the range 9 -13, a value close to that observed in CESR and NMR studies, but much higher than that observed for *ex situ* thin films and aligned crystallites.

It should be noted here that because of the high vapor pressure of Mg and its high affinity towards oxygen, it has not yet been possible to obtain *in situ* $MgB_2$ films of high chemical and structural quality with a $T_C \sim 39$ K. There are two reports of a $T_C$ (onset) over 30 K, however the corresponding films are said to be highly nanocrystalline in nature[15,16]. The observed values of $T_C(0)$ for most *in situ* films from various groups[17], including ours[18], lie in the range 22-25 K.

**Experiments**:

The $MgB_2$ films used in this study were prepared by an *in situ* approach. These c-axis oriented thin films were deposited on (1$\underline{1}$02) $Al_2O_3$ substrates using the pulsed laser deposition technique. Thin layers of Mg and B were deposited in a multilayer configuration by ablating from high purity Mg and B targets, respectively. The depositions were performed in vacuum (5 x $10^{-7}$ Torr) at room temperature, and after deposition the films were annealed at 800 $^OC$ in forming gas (4 % $H_2$ in Ar) for a short duration. The thickness of the film was about 5000 Å. The detailed description of the film deposition will be reported separately.[19]

The *R - T* measurements were performed using a standard four-probe technique in a magnetic field *H* ranging from 0 up to 8 T. The *R-T* curves were measured with *H* applied in *ab* plane (i.e. in the plane of the film) and *H* applied along the c-axis (i.e. normal to the plane).

**Results and discussions**:

Figure 1 shows the *R-T* curve for the $MgB_2$ film from 300 K down to 4.2 K. The sample shows a weak metallic behavior and the resistance drops by only a few ohms when the temperature is reduced from 300 K to 30 K. The superconducting transition starts around 27 K and below 24.5 K the sample resistance becomes zero. A rather small (1.15) residual resistance ratio, a typical signature seen in the case of *in situ* films by various groups[17], can be clearly seen from the figure. Due to small atomic numbers of Mg and B, presumably small grains in the *in situ* films, and weaker diffraction of x-rays from (00*l*) planes of MgB$_2$, well defined peaks of MgB$_2$ were not seen in the normal $\theta$-$2\theta$ x-ray diffraction (XRD) spectrum recorded at the normal scan rate. However, after performing careful slow scans on the substrates with and without the deposited films, and comparing them, a broad reflection from MgB$_2$ (002) plane became clearly visible at $2\theta = 51.45^0$. From the full width half maximum of this peak (~ $1.1^0$) and the Scherrer formula[20], an average grain size of about 8 nm could be estimated. No peaks corresponding to the grains of other orientation are observed in our XRD analysis.

Figure 2 (a) shows the *R-T* curves near the superconducting transition as a function of *H* up to 8 T applied along *c*-axis of the film. Similar data for *H* applied in *ab* plane is shown in Fig. 2(b). As expected, the superconducting transition shifts to lower temperatures with increase in the value of magnetic field. However, the decrease in $T_C$ is much less when *H* is in the *ab* plane than when it is along the *c*-axis, thus clearly revealing the highly anisotropic nature of MgB$_2$. Interestingly, in both the cases, no significant magnetoresistance is observed in the normal state for the studied range of *H*. This result is similar to the observations of Jung *et al.*[21] and Takano *et al.*[22] on high pressure sintered compound, and of Pradhan *et al.*[6] on MgB$_2$ wires synthesized by power-in-tube method, but in clear contrast to the observation of large magnetoresistance in dense MgB$_2$ wires reported by Bud'ko *et al.*[2] and in sintered bulk by Finnemore *et al.*[4]. Bud'ko *et al.*[2] have suggested that it is hard to detect the large magnetoresistance in samples with enhanced impurity or defect scattering. While the high pressure sintered materials as well as films can support residual strains and defects depending on the details of the processing conditions, the issue of the origin and control of normal state magnetoresistance remains to be resolved by further work.

An important feature noted from Fig. 2 is that the broadening of the transition occurs at lower values of *H* and further increase in *H* causes only a downward shift in temperature without any significant additional broadening of the transition. For the clear visualization of this feature, we plotted the field induced broadening in the transition, defined as $\Delta T_C(H) - \Delta T_C(0)$, where $\Delta T_C(0)$ and $\Delta T_C(H)$ are the transition widths without and with field obtained using 10-90% resistance criterion, in Fig. 3. In both the cases, most of the broadening occurs below a field of about 2 T, and at higher field only a parallel shift is noted.  Such a parallel shift of the transition has been observed in YBa$_2$Cu$_3$O$_{7-\delta}$ films deposited on MgO substrates.[23] This was interpreted to be due to a high density of defects in the film caused by large lattice mismatch between YBa$_2$Cu$_3$O$_{7-\delta}$ and MgO. Suzuki and Hikita[24] have observed in La$_{1-x}$Sr$_x$CuO$_4$ films that at lower x the transition broadens with field, however, at higher x, $T_C$ falls abruptly, showing a parallel shift of the transition. Note here that the transition broadening in magnetic field is also associated with the distribution of upper critical fields, which may arise from extrinsic sources such as the distribution in crystal orientation and presence of inhomogeneities. In this context, Takano *et al.*[22] prepared samples by sintering MgB$_2$ powder at two different temperatures, 775 and 1000 $^O$C. They observed that the superconducting transition was broader for the sample sintered at 775 $^O$C than that for the sample sintered at 1000 $^O$C. With application of magnetic field, they observed a broadening of the transition in both the samples, but the degree of broadening was much less for the sample sintered at 1000

$^O$C. Indeed, for this case, they noted a shift of the transition without enhanced broadening, as we have found in our *in situ* films. They concluded that this was due to a stronger inter-grain electrical connectivity in the 1000 $^O$C sample. It is possible that the inter-grain connectivity may be strong in our oriented films as well, and the defects responsible for the observed properties could be the intra-grain defects due to impurities or small stoichiometry departures.

From Fig. 2 we extract the information of $H_{C2}$ and $H^*$ as a function of $T$. The inset in Fig. 1(b) shows the definitions of $T^{on}$, $T^{mid}$, and $T^*$ used here to estimate $H_{C2}^{on}$, $H_{C2}^{mid}$, and $H^*$, respectively. We obtained $H_{C2}$ in two different ways: (i) from the point of maximum slope of the transition which corresponds to $H_{C2}^{mid}$, and (ii) from the point of intersection of the tangent drawn at the point of maximum slope and the line drawn through the normal state resistance data points in the vicinity of transition which corresponds to $H_{C2}^{on}$. This later definition is same as that used in Ref. 8. The $H^*$ was determined from the point of intersection of tangent drawn to the point of the steepest slope and horizontal line going through $R = 0$. This information is plotted in Fig. 4 as a function of $T$ for both the cases of $H$ along the *c*-axis and $H$ in the *ab* plane. It can be seen that these curves are quite linear except very near $T_C$. The slopes ($dH_{C2}/dT$) were extracted from the fits to the linear portions of the curves, and were used to estimate the $H_{C2}(0)$ values from the expression $H_{C2}(0) = 0.691 \times T_C \times (dH_{C2}/dT)$, given by Werthamer et al.[25] and Maki[26] for type II superconductors in the dirty limit. The values of $H_{C2}(0)$ thus obtained, are found to be in the range 100 – 130 T and about 10 T, for the field in *ab* plane ($^{ab}H_{C2}$) and along *c*-axis ($^cH_{C2}$), respectively, depending on the criterion chosen for the transition temperature. The values of $H^*$ for $H$ in *ab* plane and along *c*-axis are about 80 and 9 T, respectively.

The $^cH_{C2}$ values are similar to those obtained for bulk polycrystals[2,4,6] and aligned crystallites[13]. However, the $^{ab}H_{C2}$ values are exceptionally high. With these values, the anisotropy ratio for these films is $\eta = 9 - 13$, much higher than that observed in oriented crystallites[13] and *ex situ* thin films[7,8] but closer to that observed in single crystals[14] and indirectly obtained on polycrystalline material, separately by CESR[12] and NMR[10] techniques. The values of coherence length were determined from the following formulae[27]:

$^{ab}\xi = [\Phi_0 / (2\pi \, ^cH_{C2})]^{1/2}$    and    $^c\xi = \Phi_0 / (2\pi \, ^{ab}\xi \, ^{ab}H_{C2})$

The values obtained in the *ab* plane ($^{ab}\xi$) and along *c*-axis ($^c\xi$) are given in Table I.

In summary, we have observed a large anisotropy in the upper critical field ($H_{C2}$) for MgB$_2$ films grown *in situ* by a pulsed laser deposition process. Zero temperature values of $H_{C2}$ are found to be in the range 100 – 130 T and about 10 T, depending on the chosen criterion, for the field in *ab* plane and along *c*-axis, respectively. These values of $H_{C2}$ yield anisotropy parameter in the range 9 – 13. The estimated coherence length in *ab* plane is ~ 58 Å, whereas that in the *c*-direction is ~ 4 - 5 Å. No significant magnetoresistance is observed in the normal state. These observations suggest that in the case of *in situ* grown MgB$_2$ films, a significant concentration of defects is present, either in the form of stoichiometry defects or inclusion of impurities (most likely oxygen).

Presence of nonstoichiometric phases as well as defect phases in films could also lead to weak link effects and may contribute to anisotropy features in a complex manner. While the feature of weak broadening of the transition width with magnetic field, as observed, may be useful in certain applications, further optimization of film quality is clearly warranted.

**Acknowledgements**:
This work was supported by the Office of Naval Research under grant # N000149611026 (Program Manager: Dr. Deborah Van Vechten), and in part by NSF-MRSEC under grant # DMR-00-80008.

**Figure captions**:

Fig. 1: Resistivity versus temperature curve for $MgB_2$ film used for this study.

Fig. 2: (a) *R-T* curves near superconducting transition for *H* = 0 to 8 T applied along *c*-axis. (b) *R-T* curves near superconducting transition for *H* = 0 to 8 T applied in *ab* plane. The inset explains the definitions of $T^{on}$, $T^{mid}$, and $T^*$ used to calculate upper critical field and irreversibility field.

Fig. 3: The magnetic field induced broadening of the superconducting transition, defined as the difference between the transition widths with and without field, plotted as a function of *H*.

Fig. 4: The upper critical field and the irreversibility field plotted as a function of $T_C$ for the case of (a) *H* along *c*-axis, and (b) *H* in *ab* plane. The lines are the linear fits to the data.

Table I: The values of upper critical fields and coherence lengths along *c*-axis and in *ab* plane.

| Criterion | $^cH_{C2}(0)$ | $^{ab}H_{C2}(0)$ | $^c\xi(0)$ | $^{ab}\xi(0)$ |
|---|---|---|---|---|
| Onset | 10 T | 129 T | 4.5 Å | 57 Å |
| Maximum slope | 9.5 T | 107 T | 5.2 Å | 59 Å |

Figure 1:

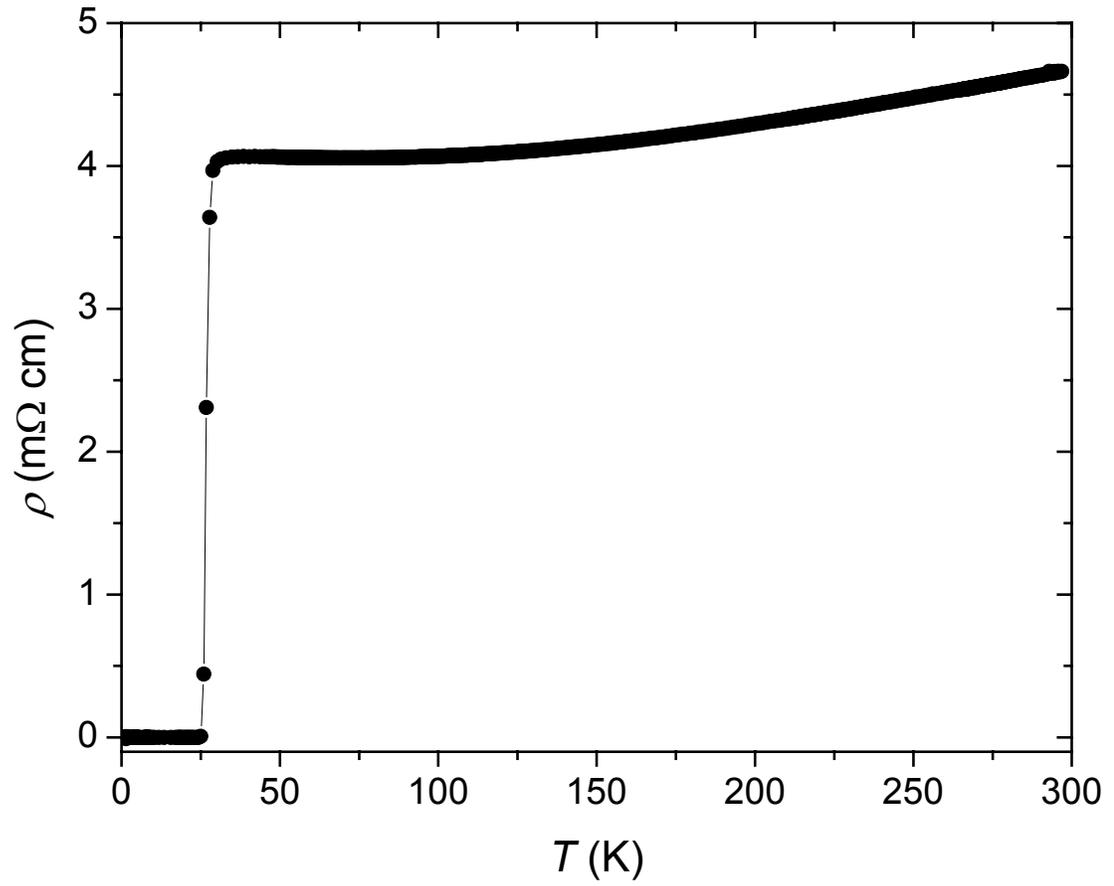

S.R. Shinde *et al*.

Figure 2:

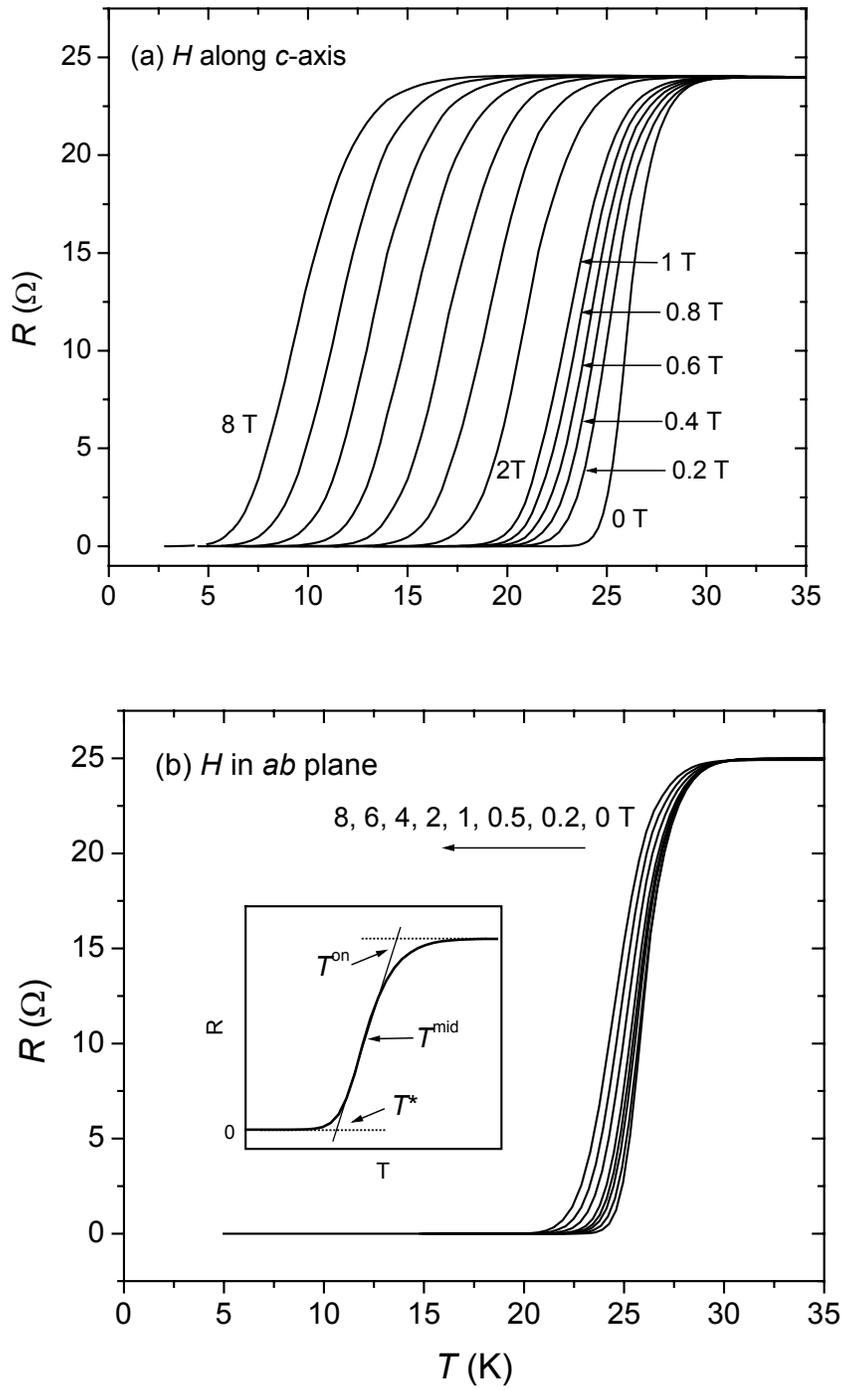

S.R. Shinde *et al*.

Figure 3:

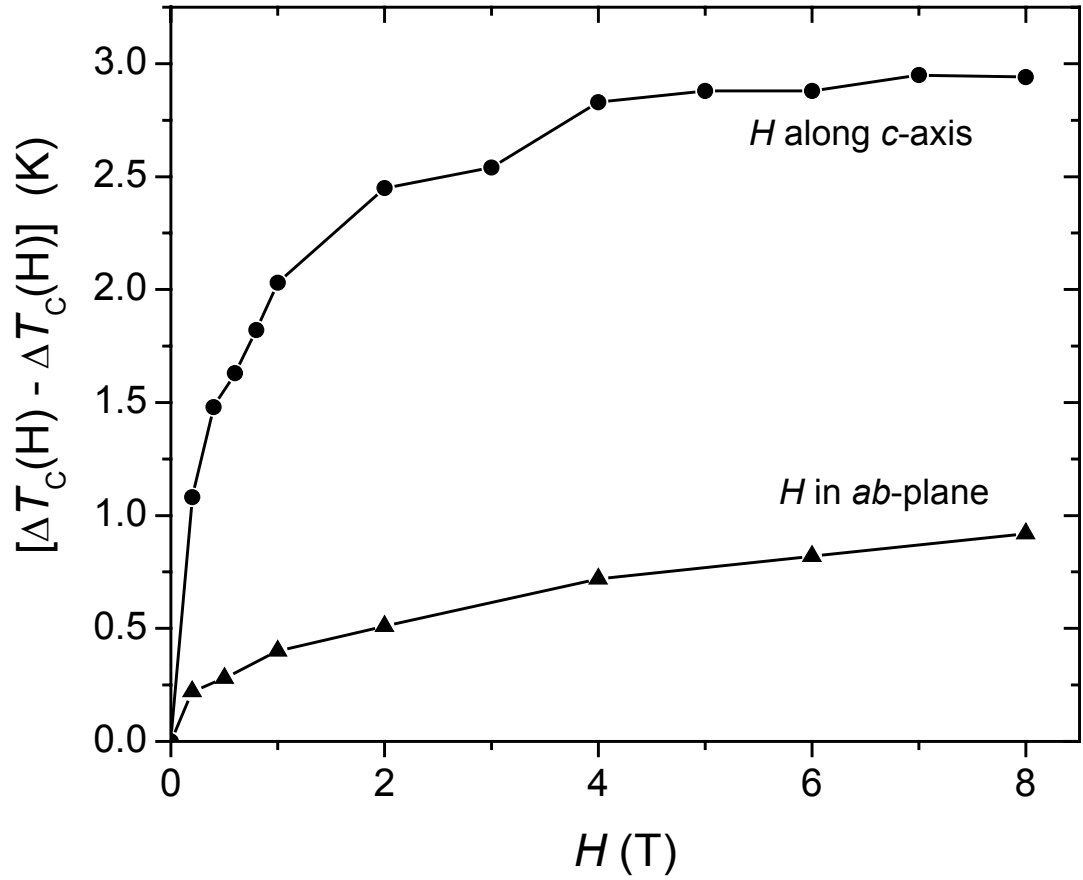

S.R. Shinde *et al*.

Figure 4:

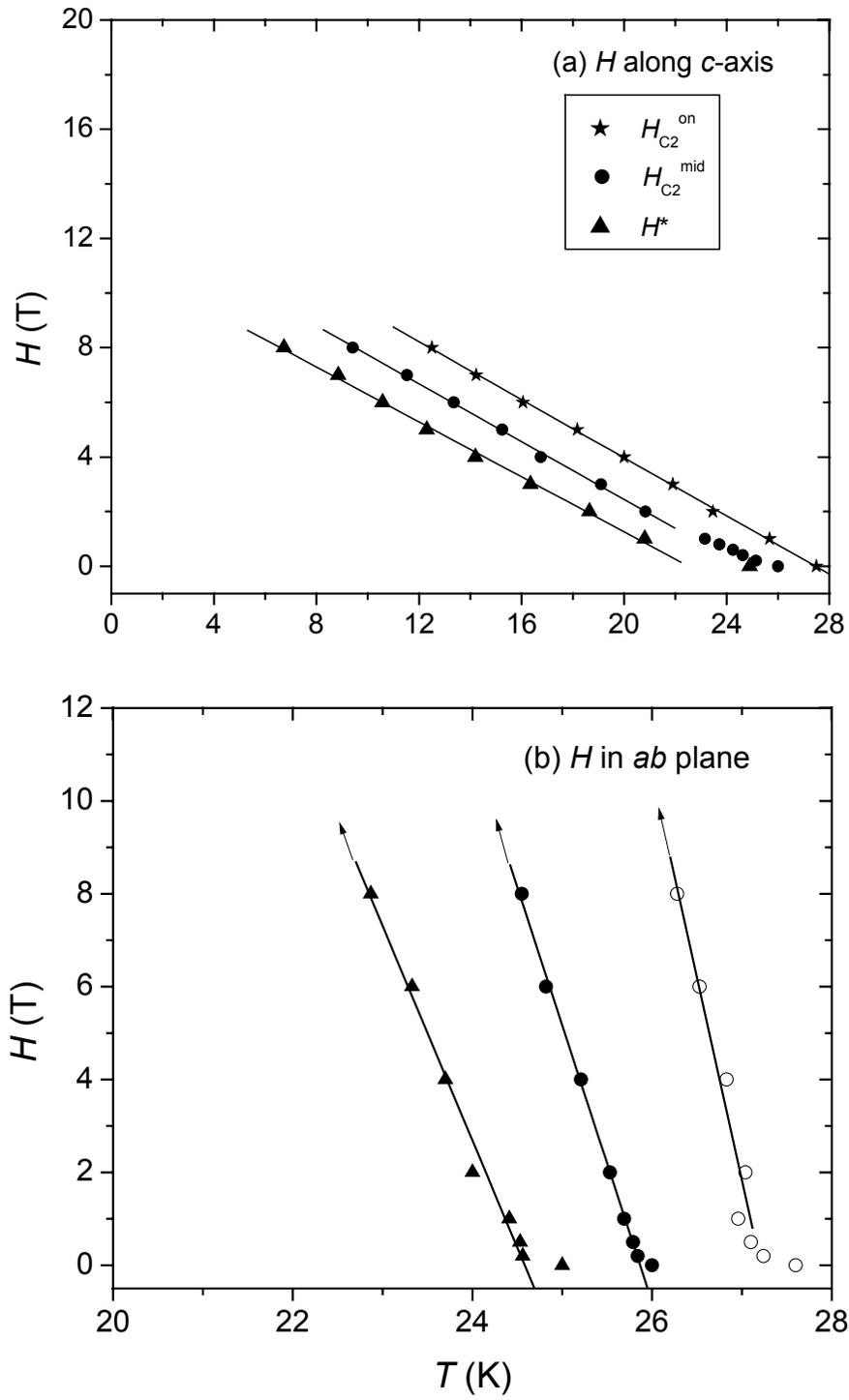

S.R. Shinde *et al*.